\def\papertitle{Wavelet Scattering on the Pitch Spiral}
\def\paperauthorA{Vincent Lostanlen}
\def\paperauthorB{St\'{e}phane Mallat}
\def\paperauthorC{Joseph Fourier}
\def\paperauthorD{Claude Shannon}
\documentclass[twoside,a4paper]{article}
\usepackage{dafx_15}
\usepackage{amsmath,amssymb,amsfonts,amsthm}
\usepackage{euscript}
\usepackage[latin1]{inputenc}
\usepackage[T1]{fontenc}
\usepackage{ifpdf}

\usepackage[english]{babel}
\usepackage{caption}
\usepackage{subfig, color}

\ninept

\usepackage{times}

\newif\ifpdf
\ifx\pdfoutput\relax
\else
   \ifcase\pdfoutput
      \pdffalse
   \else
      \pdftrue
\fi

\ifpdf 
  \usepackage[pdftex,
    pdftitle={\papertitle},
    pdfauthor={\paperauthorA, \paperauthorB, \paperauthorC, \paperauthorD},
    colorlinks=false, 
    bookmarksnumbered, 
    pdfstartview=XYZ 
  ]{hyperref}
  \usepackage[pdftex]{graphicx}
  \usepackage[figure,table]{hypcap}
\else 
  \usepackage[dvips]{epsfig,graphicx}
  \usepackage[dvips,
    colorlinks=false, 
    bookmarksnumbered, 
    pdfstartview=XYZ 
  ]{hyperref}
  \usepackage[figure,table]{hypcap}
\fi

\title{\papertitle}

\affiliation{
\paperauthorA, \paperauthorB
\sthanks{This work is supported by the ERC InvariantClass 320959. The source code to reproduce figures and experiments is available
at \texttt{www.github.com/lostanlen/scattering.m}.}}
{\href{http://di.ens.fr/data/}{Department of Computer Science},
\'{E}cole normale sup\'{e}rieure \\
Paris, France \\
{\tt \href{mailto:vincent.lostanlen@ens.fr}{vincent.lostanlen@ens.fr}}
}
\usepackage{amsopn}
\DeclareMathOperator{\sign}{sign}

\begin{document}

\ifpdf 
  \DeclareGraphicsExtensions{.png,.jpg,.pdf}
\else  
  \DeclareGraphicsExtensions{.eps}
\fi

\maketitle

\begin{abstract}
We present a new representation of harmonic sounds that linearizes the dynamics of pitch and spectral envelope, while remaining stable to deformations in the time-frequency plane. It is an instance of the scattering transform, a generic operator which cascades wavelet convolutions and modulus nonlinearities. It is derived from the pitch spiral, in that convolutions are successively performed in time, log-frequency, and octave index. We give a closed-form approximation of spiral scattering coefficients for a nonstationary generalization of the harmonic source-filter model.
\end{abstract}

\section{Introduction}

The spectro-temporal evolution of harmonic sounds conveys essential information to audio classification, blind source separation, transcription, as well as other processing tasks. This information is however difficult to capture in time-varying, polyphonic mixtures.
On one hand, spectrogram-based pattern recognition algorithms \cite{Kereliuk2008} are exposed to detection errors as they enforce strong constraints on the shape of harmonic templates. On the other, time-varying generalizations of matrix factorization \cite{Hennequin2011} are under-constrained and thus may fail to converge to a satisfying solution. In this article, we address the characterization of harmonic structures without any detection nor training step.

Wavelets have long proven to provide meaningful, sparse activations as long as they operate on a dimension on which the signal has already some regularity. Although a single sine wave draws a regular edge on the time-frequency plane, a harmonic comb is made of distant sharp peaks over the log-frequency axis, an irregular pattern that is hard to characterize globally. This irregularity weakens the discriminative power of existing wavelet-like representations, such as Mel-frequency cepstral coefficients (MFCC).

To recover regularity across partials within a wavelet framework, we capitalize on the fact that power-of-two harmonics are exactly one octave apart. By rolling up the log-frequency axis into a spiral, such that octave intervals correspond to full turns, these partials get aligned on a radius. Consequently, introducing the integer-valued octave variable reveals harmonic regularity that was not explicit in the plane of time and log-frequency.

Once specified the variables of time, log-frequency, and octave index, our representation merely consists in cascading three wavelet decompositions along them and applying complex modulus. Thus, the constant-Q scalogram is "scattered" into channels over which main factors of time variability are disentangled and regularized, yet harmonicity is preserved.

Section 2 gives a formal definition of the spiral scattering transform. Section 3 introduces a nonstationary formulation of the source-filter model relying on time warps, and shows that its variabilities in pitch and spectral envelope are jointly linearized by the spiral scattering transform. Section 4 provides a visual interpretation of the spiral scattering coefficients of a nonstationary musical note.

\begin{figure}[t]
    \begin{center}
        \setlength{\unitlength}{1cm}
        \begin{picture}(8.5,5.5)
        \put(0,0){\includegraphics[width=8.5cm]{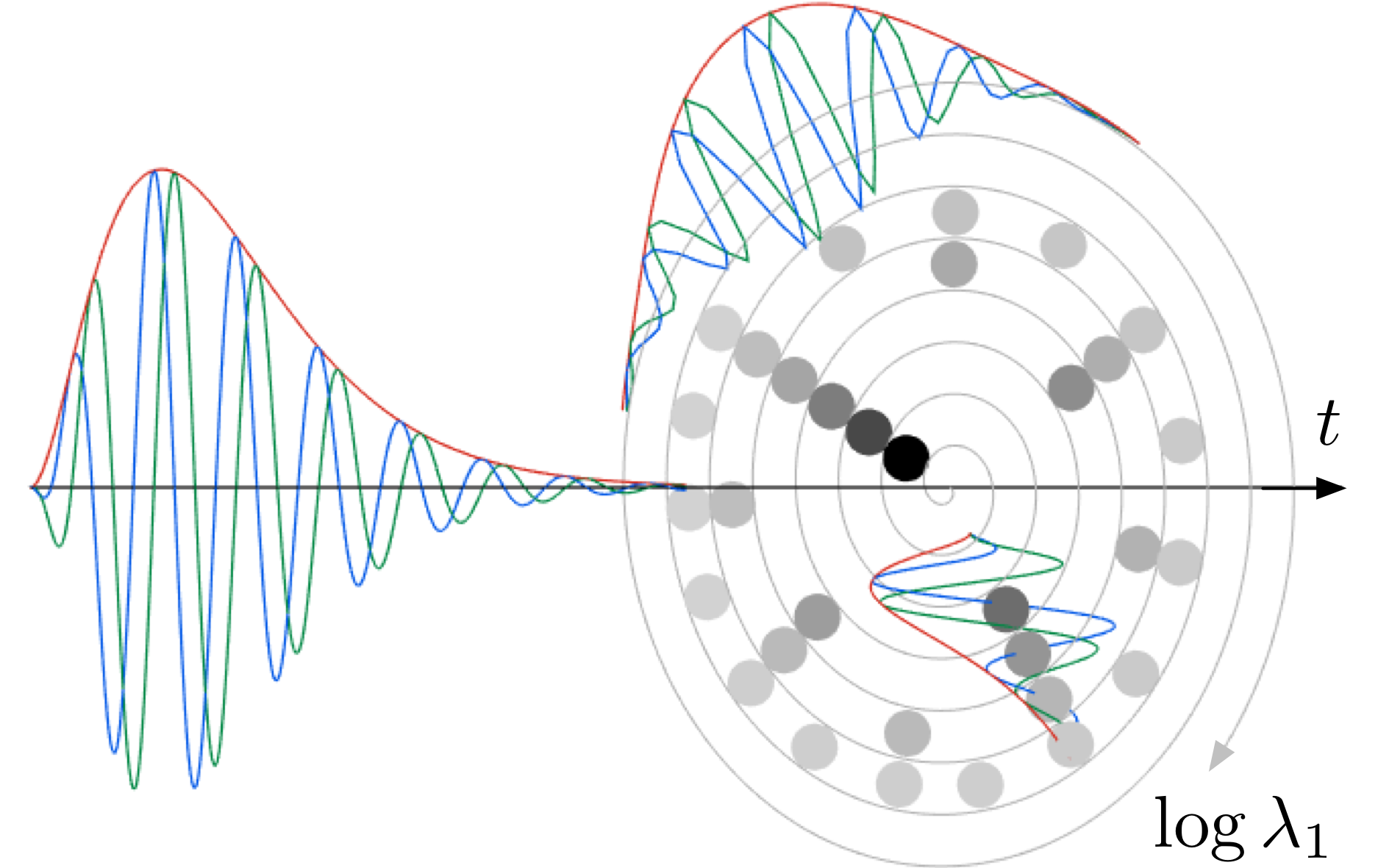}}
        \end{picture}
    \end{center}
\protect\caption{
The spiral wavelet is a product of wavelets along time, log-frequency, and octave index. Blue and green oscillations represent the real and imaginary parts. The red envelope represents the complex modulus.
Partials of an hypothetical harmonic sound are marked as thick dots.
\label{fig:spiral-wavelets}
}
\end{figure}

\section{From time scattering to spiral scattering}
This section builds the spiral scattering transform progressively as a cascade of wavelet transforms along time, log-frequency, and octave index. All three variables share the same framework.

\subsection{Time scattering}

An analytic "mother" wavelet is a complex filter $\psi(t)$
whose Fou\-rier transform $\widehat{\psi}(\omega)$ is concentrated over the
dimensionless frequency interval $[1-Q/2,1+Q/2]$, where the quality factor $Q$ is in the typical range $12$\textendash$24$. Dilations of this wavelet
define a family of bandpass filters centered at frequencies
$ \lambda_{1} = 2^{j_{1} + \frac{\chi_{1}}{Q}}$,
where the indices $j_{1} \in \mathbb{Z}$ and $\chi_{1} \in \{1\ldots\,Q\}$ respectively denote octave and chroma:
\begin{equation}
    \widehat{\psi}_{\lambda_{1}}(\omega) = \widehat{\psi}(\lambda^{-1}\omega)
    \quad\mathrm{i.e.}\quad
    \psi_{\lambda_{1}}(t)=\lambda_{1}\psi(\lambda_{1}t).
\label{eq:wavelet-dilations}
\end{equation}

The wavelet transform convolves an input signal $x(t)$ with the filter bank of $\psi_{\lambda_1}$'s. We denote convolutions along time by the operator $\overset{t}{\ast}$. Applying complex modulus to all wavelet convolutions results in the "scalogram" matrix
\begin{equation}
    x_1(t, \log \lambda_1) =
    |x \overset{t}{\ast} \psi_{\lambda_1}| \quad \text{for all } \lambda_1 > 0,
\label{eq:scalogram}
\end{equation}
whose frequential axis is uniformly sampled by the binary logarithm $\log \lambda_1$. The scalogram $x_1$ localizes the energy of $x(t)$ around frequencies $\lambda_1$ over durations $2 Q / \lambda_1$, trading frequency resolution for time resolution.

The constant-Q transform (CQT) $S_1 x$ corresponds to a lowpass filtering of $x_1$ with a window $\phi_T(t)$ of size $T$:
\begin{equation}
S_1 x (t, \log \lambda_1) =
x_1 \overset{t}{\ast} \phi_T =
| x \overset{t}{\ast} \psi_{\lambda_{1}} | \overset{t}{\ast} \phi_T.
\label{eq:S1}
\end{equation}

To recover the amplitude modulations lost when averaging by $\phi_T$ in
Equation (\ref{eq:S1}), the time scattering transform also convolves $x_1$ with a second filterbank of wavelets $\psi_{\lambda_2}$ and applies complex modulus to get
\begin{equation}
x_2 (t, \log \lambda_1, \log \lambda_2) =
\vert x_1 \overset{t}{\ast} \psi_{\lambda_2} \vert =
\vert \vert x \overset{t}{\ast} \psi_{\lambda_{1}} \vert
\overset{t}{\ast} \psi_{\lambda_{2}} \vert.
\label{eq:x2-time}
\end{equation}

The wavelets $\psi_{\lambda_2}(t)$ have a quality factor in the range $1$\textendash$2$, though we choose to keep the same notation $\psi$ for simplicity. Like in Equation (\ref{eq:S1}), averaging in time creates invariance to translation in time up to $T$, yielding
\begin{equation}
S_2 x (t, \log \lambda_1, \log \lambda_2) =
x_2 \overset{t}{\ast} \phi_T =
\vert \vert x \overset{t}{\ast} \psi_{\lambda_{1}} \vert
\overset{t}{\ast} \psi_{\lambda_{2}} \vert
\overset{t}{\ast} \phi_T.
\label{eq:S2-time}
\end{equation}

Due to the constant-Q property, $S_1 x$ and $S_2 x$ are stable to small time warps of $x(t)$ as long as they do not exceed $Q^{-1}$, i.e. one semitone. This implies that small modulations, such as tremolo and vibrato, are accurately linearized \cite{Anden2012}.

\subsection{Joint time-frequency scattering}

The time scattering transform defined in Equation (\ref{eq:x2-time}) decomposes each frequency band separately, and thus cannot properly capture the coherence of time-frequency structures, such as those induced by pitch contour. To remedy this, And\'{e}n et al. \cite{Anden2015} have redefined the wavelets $\psi_{\lambda_2}$'s as functions of both time and log-frequency, indexed by pairs $\lambda_2 = (\alpha,\beta)$, where $\alpha$ is a modulation frequency in Hertz and $\beta$ is a frequency along log-frequencies in cycles per octaves. The joint wavelets $\psi_{\lambda_2}(t,\log \lambda_1)$ factorize as
\begin{equation}
\psi_{\lambda_2}(t,\log \lambda_1) = \psi_\alpha (t) \times \psi_\beta (\log \lambda_1).
\label{eq:wavelet-joint}
\end{equation}

We write $\overset{\chi_{1}}{\ast}$ to denote convolutions along the log-frequency axis, i.e. along chromas.  Wavelet scattering is extended to two-dimensional convolutions by plugging Equation (\ref{eq:wavelet-joint}) into the definition of $x_2$ in Equation (\ref{eq:x2-time}):
\begin{equation}
x_{2}(t,\log \lambda_{1},\log \lambda_{2}) =
\vert x_{1} \overset{t, \chi_{1}}{\ast} \psi_{\lambda_{2}} \vert =
\vert x_{1} \overset{t}{\ast} \psi_{\alpha} \overset{\chi_{1}}{\ast} \psi_{\beta} \vert.
\label{eq:x2-joint}
\end{equation}

The joint time-frequency scattering transform corresponds to the "cortical transform" introduced by Shamma and his team to formalize his findings in auditory neuroscience \cite{Patil2012}.

\subsection{Spiral scattering}

The time-frequency scattering transform defined in Equation (\ref{eq:x2-joint}) provides template-free features for pitch variability along time. However, it is unaware of the harmonic structure of voiced sounds, such as vowels or musical notes. The multiscale evolution of this structure yields relevant information about attack transients and formantic changes, almost independently from the pitch contour.

In order to capture this information, we extend the joint time-frequency scattering transform to encompass regularity in time across octaves at fixed chroma, in conjunction with regularity along neighboring constant-Q bands. Just like wavelet filterbanks along time and log-frequency have been defined in the two previous subsections, we capitalize on harmonicity by introducing a third wavelet filterbank.

We roll up the log-frequency variable $\log \lambda_1$ into a pitch spiral making one full turn at each octave (see Figure \ref{fig:spiral-wavelets}). Since a frequency interval of one octave corresponds to one unit in binary logarithms $\log \lambda_1$, pitch height and pitch chroma in the spiral correspond to integer part $\lfloor \log \lambda_1 \rfloor$ and fractional part $\{ \log \lambda_1 \}$:

\begin{equation}
\log \lambda_1 = \lfloor \log \lambda_1 \rfloor + \{ \log \lambda_1 \} = j_1 + \frac{\chi_1}{Q}.
\label{eq:integer-part and fractional part}
\end{equation}

In this setting, the fundamental frequency $f_0$ is aligned with its power-of-two harmonics $2 f_0$, $4 f_0$, $8 f_0$ and so forth. Likewise, the perfect fifth $3 f_0$ is aligned with $6 f_0$. As the number of harmonics per octave increase exponentially, the alignment of upper harmonics \textemdash{} $5 f_0$, $7 f_0$, and so forth \textemdash{} in the spiral is less crucial, because it can also be recovered with convolutions along chromas for $\beta^{-1}$ of the order of a few semitones.

\begin{figure}[t]
    \begin{center}
        \setlength{\unitlength}{1cm}
        \begin{picture}(8,7)
        \put(0,0){\includegraphics[height=7cm]{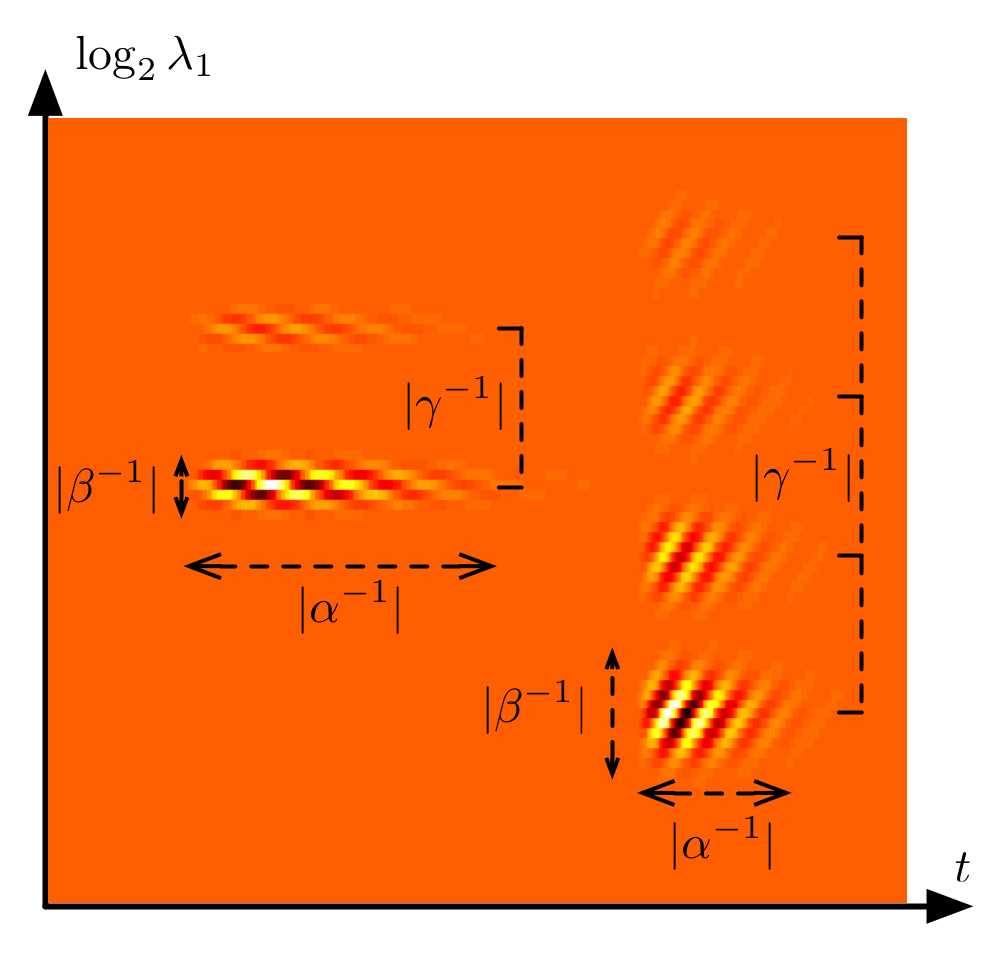}}
        \end{picture}
    \end{center}
    \protect\caption{
    Two spiral wavelets $\psi_{\lambda_2}(t,\log \lambda_1)$ in the time-frequency plane, with different values of $\lambda_2 = (\alpha,\beta,\gamma)$. Left: $\alpha^{-1} = 120\text{ ms}$, $\beta^{-1} = - 0.25\text{ octave}$, $\gamma^{-1} = +2\text{ octaves}$. Right: $\alpha^{-1} = 60\text{ ms}$, $\beta^{-1} = +0.5\text{ octave}$,  $\gamma^{-1} = -4\text{ octaves}$. Darker color levels corresponds to greater values of the real part.
\label{fig:unrolled-spiral-wavelets}
}
\end{figure}

The wavelet $\psi_{\lambda_2}$ is now defined as a product between wavelets in time, log-frequency, and octave index:
\begin{equation}
\psi_{\lambda_2}(t, \log \lambda_1) =
\psi_{\alpha}(t) \times
\psi_{\beta}(\log \lambda_1) \times
\psi_{\gamma}(\lfloor \log \lambda_1 \rfloor).
\label{eq:wavelet-shepard}
\end{equation}

In the previous definition, $\psi_{\beta}$ operates over the rectilinear log-frequency variable $\log \lambda_1$, not over the circular chroma variable $\chi_1 = \{ \log \lambda_1 \}$. This is to avoid artifacts at octave boundaries.

Examples of the "spiral wavelet" $\psi_{\lambda_2}$ are shown in Figure \ref{fig:unrolled-spiral-wavelets} for different values of $\alpha$, $\beta$ and $\gamma$. To ensure invertibility and energy conservation, the quefrencies $\beta$ and $\gamma$ must take negative values, including zero. We adopt the shorthand notation
\begin{equation}
\log \lambda_2 =
\left( \log \alpha,
\log \vert \beta \vert, \sign \beta,
\log \vert \gamma \vert, \sign \gamma \right)
\end{equation}
to specify their indexing. In the special case $\beta = 0$, $\psi_\beta$ is no longer a wavelet but a low-pass filter whose support covers one octave. By convention, the corresponding log-frequency index is $\log \vert \beta \vert = - \infty$. The same remark applies to $\psi_{\gamma}$ for $\gamma=0$, which covers six octaves. Since its Fourier transform $\widehat{\psi}_{\lambda_2}$  is centered at $(\alpha,\beta,\gamma)$, the spiral wavelet $\psi_{\lambda_2}$ has a pitch chroma velocity of $\alpha/\beta$ and a pitch height velocity of $\alpha/\gamma$, both measured in octaves per second.

We write $\overset{j_{1}}{\ast}$ to denote convolutions across neighboring octaves.
The definition for $x_2$ is comparable to Equations (\ref{eq:x2-time}) and (\ref{eq:x2-joint}):
\begin{equation}
\begin{split}
x_{2}(t,\log \lambda_{1},\log \lambda_{2})
& = \vert x_{1} \overset{t,\chi_{1},j_{1}}{\ast} \psi_{\lambda_{2}}\vert \\
&= \vert x_{1} \overset{t}{\ast} \psi_\alpha \overset{\chi_{1}}{\ast} \psi_\beta \overset{j_{1}}{\ast} \psi_\gamma \vert
.
\end{split}
\label{eq:x2-spiral}
\end{equation}

Rolling up pitches into a spiral is a well-established idea in music, if only because of circularity of musical pitch classes. It has been studied by Shepard \cite{Shepard1964}, Risset \cite{Risset1969}, and Deutsch \cite{Deutsch2008} to build paradoxes in perception of pitch, and is corroborated by functional imaging of the auditory cortex (see Warren \cite{Warren2003}).

\section{Deformations of the source-filter model}
A classical model for voiced speech production consists in the convolution of a harmonic glottal source $e(t)$ with a vocal tract filter $h(t)$.
Introducing independent deformations to both components brings realistic variability to pitch and spectral envelope.
This section studies the decomposition of the deformed source-filter model in the spiral scattering transform.

\subsection{Overview}

Let $e(t)=\sum_{n}\delta(t - n)$ be a harmonic signal and
$t\mapsto\theta(t)$ a time warp function. We define a warped source
as $e_{\theta}(t)=(e\circ\theta)(t)$. Similarly, we compose a
filter $h(t)$ and a warp $t\mapsto\eta(t)$ to define $h_{\eta}(t)=(h\circ\eta)(t)$.
The warped source-filter model is the signal
\begin{equation}
x_{\theta, \eta}(t) =
(e_{\theta} \overset{t}{\ast} h_{\eta})(t)
\label{eq:source-filter}
\end{equation}
Observe that $\dot{\theta}(t)$ induces a change of fundamental frequency, where\-as
$\dot{\eta}(t)$ accounts for a local dilation of the spectral
envelope $\vert\widehat{h}\vert(\omega)$.
We show in this section that, for $\dot{\theta}(t)$ and $\dot{\eta}(t)$ reasonably regular over the support of first-order wavelets,
the local maxima of $x_{2}$ are clustered on a plane in the $(\alpha,\beta,\gamma)$ space of scattering coefficients.
This plane satisfies the Cartesian equation
\begin{equation}
\alpha +
\dfrac{\ddot{\theta}(t)}{\dot{\theta}(t)}\beta +
\dfrac{\ddot{\eta}(t)}{\dot{\eta}(t)}\gamma = 0.
\label{eq:cartesian-equation}
\end{equation}
In a polyphonic context, this result means that harmonic sounds overlapping both in time and frequency could be resolved according to their respective source-filter velocities.

Our proof is driven by harmonicity and spectral smoothness properties \textemdash{} Equation (\ref{eq:zero-mean}) \textemdash{} and derives Equation (\ref{eq:cartesian-equation}) from the computation of wavelet ridges on the pitch spiral \cite{Delprat1992}.

\subsection{Source-filter factorization in the scalogram}

Given $\lambda_1$ near the $p^{\textrm{th}}$ partial $p \dot{\theta}(t)$ where $p \in \mathbb{N}$, we linearize $\theta(t)$ and $\eta(t)$ over the support of the first-order wavelet $\psi_{\lambda_1}(t)$. We work under the following assumptions:
\renewcommand{\labelenumi}{(\alph{enumi})}
\begin{enumerate}
\item $Q$ large enough to discriminate the $p^\text{th}$ partial: $Q>2p$,
\item slowly varying source: $\Vert\ddot{\theta}/\dot{\theta}\Vert_{\infty} \ll \lambda_{1}/Q$,
\item slowly varying filter: $\Vert\ddot{\eta}/\dot{\eta}\Vert_{\infty}\ll\lambda_{1}/Q$, and
\item spectral smoothness: \\
$\hphantom{\quad} \Vert\mathrm{d}( \log \vert \hat{h} \vert ) / \mathrm{d} \omega \Vert_{\infty} \times \Vert 1 / \dot{\eta} \Vert_{\infty} \ll Q / \lambda_{1}$.
\end{enumerate}
According to (a), partials $p^{\prime} \neq p$ have a negligible contribution to the scalogram of the source at the log-frequency $\log \lambda_1$. For lack of any interference, this scalogram is constant through time, and we may drop the dependency in $t$:
\begin{equation}
\vert e \overset{t}{\ast} \psi_{\lambda_1} \vert \approx
\vert \widehat{\psi}_{\lambda_1}(p) \vert.
\end{equation}
According to (b), the scalogram of the warped source $e_\theta (t)$ can be replaced by the scalogram of the original source translated along the log-frequency axis at the velocity $\log \dot{\theta}(t)$:
\begin{equation}
\vert e_{\theta} \overset{t}{\ast} \psi_{\lambda_1} \vert (t) =
\vert e \overset{t}{\ast} \psi_{\lambda_1} \vert (\theta(t)) \approx
\vert \widehat{\psi}_{\lambda_1}(p \dot{\theta}(t)) \vert.
\label{eq:e-theta-1}
\end{equation}
According to (c), we linearize $\eta(t)$ over the support of $\psi_{\lambda_1}(t)$. According to (d), we approximate $\hat{h}(\omega)$ by a constant over the frequential support of the wavelet and factorize the filtering as a product:
\begin{equation}
\left( h_\eta \ast \psi_{\lambda_1} \right)(t) \approx
\hat{h}\left(\frac{\lambda_1}{\dot{\eta}(t)}\right)
\times
\psi_{\lambda_1}\left( \frac{\eta(t)}{\dot{\eta}(t)} \right).
\label{eq:h-nu-1}
\end{equation}
By plugging Equation (\ref{eq:e-theta-1}) into Equation (\ref{eq:h-nu-1}), the scalogram of the deformable source-filter model appears as a separable product:
\begin{equation}
\vert x_{\theta,\eta} \overset{t}{\ast} \psi_{\lambda_1} \vert (t) =
\vert \widehat{\psi}_{\lambda_1}(p \dot{\theta}(t)) \vert
\times
\left \vert \hat{h}\left(\frac{\lambda_1}{\dot{\eta}(t)}\right) \right \vert.
\label{eq:factorization}
\end{equation}

\subsection{Harmonicity and spectral smoothness properties}

The second step in the proof consists in showing that the convolution along chromas with $\psi_\beta$ only applies to $e_{1,\theta}$, whereas the convolution across octaves with $\psi_\gamma$ only applies to $h_{1,\eta}$. Indeed, all wavelets are designed to carry a negligible mean value, i.e. convolving them with a constant yields zero. Therefore, the harmonicity and spectral smothness properties rewrite as
\begin{equation}
\Big\vert \vert e_\theta \overset{t}{\ast} \psi_{\lambda_1} \vert \overset{j_1}{\ast} \psi_{\gamma} \Big\vert \approx 0
\quad
\text{ and }
\quad
\Big\vert \vert e_\theta \overset{t}{\ast} \psi_{\lambda_1} \vert \overset{\chi_1}{\ast} \psi_{\beta} \Big \vert \approx 0.
\label{eq:zero-mean}
\end{equation}
Gathering Equations (\ref{eq:factorization}) and (\ref{eq:zero-mean}) into the definition of spiral scattering yields
\begin{eqnarray}
x_{\theta,\eta}\overset{t,\chi_{1},j_{1}}{\ast}\psi_{\lambda_{2}}  \mkern-72mu \\
 = & \left[\Big(\vert e_{\theta}\overset{t}{\ast}\psi_{\lambda_{1}}\vert\overset{\chi_{1}}{\ast}\psi_{\beta}\Big)\times \Big(\vert h_{\eta}\overset{t}{\ast}\psi_{\lambda_{1}}\vert\overset{j_{1}}{\ast}\psi_{\gamma}\Big)\right]\overset{t}{\ast}\psi_{\alpha} \nonumber,
\end{eqnarray}
where the superscripts $t$, $\chi_{1}$, and $j_{1}$ denote convolutions along time, chromas and octaves respectively.

\subsection{Extraction of instantaneous frequencies}
As a final step, we state that the phase of $( \vert e_\theta \overset{t}{\ast} \psi_{\lambda_1} \vert \overset{\chi}{\ast} \psi_{\beta} )$ is $\beta\times(\log \lambda_1 - \log p \dot{\theta}(t))$. By differentiating this quantity along $t$ for fixed $\log \lambda_1$, we obtain an instantaneous frequency of $- \beta \ddot{\theta}(t)/\dot{\theta}(t)$. Similarly, the instantaneous frequency of $( \vert h_\eta \overset{t}{\ast} \psi_{\lambda_1} \vert \overset{j_{1}}{\ast} \psi_{\gamma} )$ is $ - \gamma \ddot{\eta}(t)/\dot{\eta}(t)$. As long as
\begin{equation}
\alpha \geq \left \vert \frac{\ddot{\theta}(t)}{\dot{\theta}(t)} \beta \right \vert
\quad
\text{and}
\quad
\alpha \geq \left \vert \frac{\ddot{\eta}(t)}{\dot{\eta}(t)} \gamma \right \vert,
\end{equation}
the envelopes of these two convolutions are almost constant over the support of $\psi_{\alpha}(t)$ \cite{Delprat1992}. We conclude with the following approximate closed-form expression for the spiral scattering coefficients of the deformed source-filter model:
\begin{eqnarray}
\mkern-60mu
x_2(t,\log \lambda_1, \log \lambda_2) = 
\Big\vert \vert e_\theta \overset{t}{\ast} \psi_{\lambda_1} \vert \overset{\chi_1}{\ast} \psi_{\beta} \Big\vert
\times
\Big\vert \vert h_\eta \overset{t}{\ast} \psi_{\lambda_1} \vert \overset{j_1}{\ast} \psi_{\gamma} \Big\vert \nonumber \mkern-310mu \\
& \times \left \vert \widehat{\psi}_{\alpha} \left( -\dfrac{\ddot{\theta}(t)}{\dot{\theta}(t)} \beta - \dfrac{\ddot{\eta}(t)}{\dot{\eta}(t)} \gamma \right) \right \vert.
\label{eq:x2-sourcefilter}
\end{eqnarray}
The Fourier spectrum $\vert \widehat{\psi}_\alpha(\omega) \vert$ of $\psi_\alpha(t)$ is a bump centered at the frequency $\alpha$. Equation (\ref{eq:cartesian-equation}) follows immediately from the above formula. The same result holds for the averaged coefficients $S_2 x = x_2 \ast \phi_T$ if the velocities $\ddot{\theta}(t)/\dot{\theta}(t)$ or $\ddot{\eta}(t)/\dot{\eta}(t)$ have small relative variations:
\begin{equation}
\left \vert \frac{\dddot{\theta}(t)}{\ddot{\theta}(t)} - \frac{\ddot{\theta}(t)}{\dot{\theta}(t)} \right \vert \ll T^{-1}
\quad \text{and} \quad
\left \vert \frac{\dddot{\eta}(t)}{\ddot{\eta}(t)} - \frac{\ddot{\eta}(t)}{\dot{\eta}(t)} \right \vert \ll T^{-1}.
\end{equation}
\begin{figure}[t]
    \begin{center}
        \setlength{\unitlength}{1cm}
        \begin{picture}(8,10.5)
        \put(0,0){\includegraphics[width=8cm]{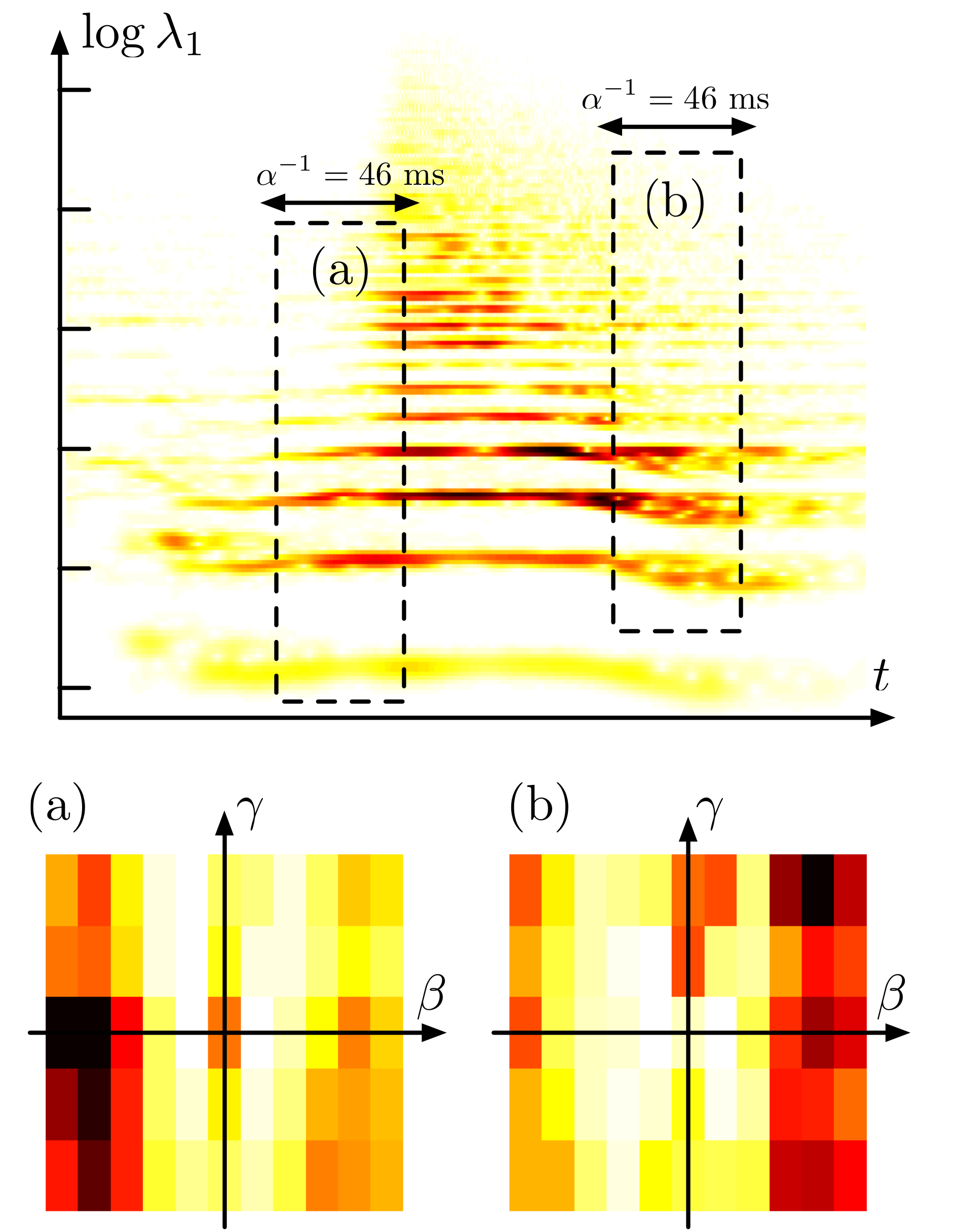}}
        \end{picture}
    \end{center}
    \protect\caption{
    Top: scalogram of a musical note from Berio's \emph{Sequenza V} for trombone, around 3'45".
    Observe that the attack part (a) has increasing pitch and increasing brightness, whereas the release part (b) has decreasing pitch and decreasing brightness.
    Bottom: spiral scattering coefficients for $t$ and $\log \lambda_1$ specified by (a) and (b), $\alpha^{-1}$ fixed at 46 ms, $\beta^{-1}$ ranging from $-1\text{ octave}$ to $+1\text{ octave}$, and $\gamma^{-1}$ ranging from $-4\text{ octaves}$ to $+1\text{ octaves}$. As expected, highest values are concentrated in the bottom left corner for (a) and in the top right corner for (b).
\label{fig:berio-scalogram}
}
\end{figure}
In the example of a trombone signal, glissando can be modeled by $\ddot{\theta}(t) / \dot{\theta}(t)$ in the source-filter model, whereas the brassiness profile induces a timbral velocity $\ddot{\eta}(t) / \dot{\eta}(t)$. Figure 3 illustrates that these two velocities are stably disentangled and characterized.

\section{Conclusions}

The spiral model we have presented is well-known in music theory and experimental psychology \cite{Shepard1964, Risset1969, Deutsch2008}. However, existing methods in audio signal processing do not fully take advantage from its richness, because they either picture pitch on a line (e.g. MFCC) or on a circle (e.g. chroma features). In this article, we have shown how spiral scattering can represent the transientness of harmonic sounds.

\nocite{*}
\bibliographystyle{IEEEbib}
\bibliography{Lostanlen_DAFx15} 

\end{document}